# An Aligned Stream of Low-Metallicity Clusters in the Halo of the Milky Way

Suk-Jin Yoon[*] and Young-Wook Lee

Center for Space Astrophysics, Yonsei University, Seoul 120–749, Korea

[*] To whom correspondence should be addressed. E-mail: sjyoon@csa.yonsei.ac.kr

**ABSTRACT**

**One of the long-standing problems in modern astronomy is the curious division of Galactic globular clusters, the "Oosterhoff dichotomy," according to the properties of their RR Lyrae stars. Here, we find that most of the lowest-metallicity ([Fe/H]<-2.0) clusters, which are essential to an understanding of this phenomenon, display a planar alignment in the outer halo. This alignment, combined with evidence from kinematics and stellar population, indicates a captured origin from a satellite galaxy. We show that, together with the horizontal-branch evolutionary effect, the factor producing the dichotomy could be a small time-gap between the cluster-formation epochs in the Milky Way and the satellite. The results oppose the traditional view that the metal-poorest clusters represent the indigenous and oldest population of the Galaxy.**

More than 60 years ago, Oosterhoff (*1*) discovered that Galactic globular clusters could be divided into two distinct groups according to the mean period of type ab RR Lyrae variables ($\langle P_{ab} \rangle$). This dichotomy was one of the earliest indications of systematic difference among globular clusters, whose reality has been strengthened by subsequent investigations (*2*). Given that most characteristics of Galactic globular clusters appear to be distributed in a continuous way, it is unusual that a quantity used to characterize variable stars, falls into two rather well-defined classes. Moreover, the two groups are known to differ in metal abundance (*3*) and kinematic properties (*4*), which may indicate distinct origins. Whatever the reasons for the dichotomy, the question of whether the two groups originated under fundamentally different conditions is of considerable interest regarding the formation scenarios of the Galactic halo. Despite many efforts during the last decades, the origin of this phenomenon still lacks a convincing explanation.

Figure 1 shows the Oosterhoff dichotomy. Clusters belong to groups I and II if their values of $\langle P_{ab} \rangle$ fall near 0.55 and 0.65 days, respectively (*5*). Group I is more metal-rich than group II (*3*). Based on our horizontal-branch (HB) population models (*6, 7*), we have found that the presence of the relatively metal-rich (-1.9<[Fe/H]<-1.6) clusters in group II (hereafter group II-a) can be understood by the sudden increase in $\langle P_{ab} \rangle$ at [Fe/H]≈-1.6 (indicated by a blue line in Fig. 1). As [Fe/H] decreases, HB stars get hotter (i.e., the HB morphology gets bluer), and there is a certain point at which the zero-age portion of the HB just crosses the blue edge of the instability strip (Fig. 2). Then, only highly evolved stars from the blue side of the instability strip can penetrate back into the strip, and thus the mean RR Lyrae luminosity and $\langle P_{ab} \rangle$ increase abruptly.

However, this effect alone cannot explain the observed periods of the relatively metal-poor ([Fe/H]<-2.0) clusters in group II (hereafter group II-b) because $\langle P_{ab} \rangle$ continues to increase as the HB slides monotonically to blue with decreasing [Fe/H] (as indicated by the blue line in Fig. 1). In reality, the HB morphologies of group II-b clusters are redder than predicted by the models at their metallicity, implying that the general trend of the HB to become bluer with decreasing metallicity is reversed among the most metal-poor clusters (*8*). We believe that this nonmonotonic behavior of HB morphology is the key to the complete understanding of the Oosterhoff dichotomy. Here, we find from the HB models that, if the mean age of group II-b is slightly younger [Δage≈1 billion years (Gy)] than that of group II-a, then the observed HB morphology and $\langle P_{ab} \rangle$ of group II-b can be reproduced (Fig. 1). This is because, in younger clusters, HB morphology is redder at a given [Fe/H] than it otherwise would be (*7*), and the redward evolution from blue HB comes into effect at more metal-poor region ([Fe/H]≈-2.0).

Although the required age difference is small and comparable to the current uncertainty in relative age dating of globular clusters, this solution contradicts the general notion that the most metal-poor clusters like group II-b represent the components formed during the initial collapse of the proto-Galaxy (*9*) and hence are among the oldest objects in the Milky Way (*10*). There is now a growing body of evidence that satellite galaxies have contributed their globular clusters to the Milky Way halo (*11*), but there has hitherto been no evidence that the lowest-metallicity clusters also have such an origin. As described below, however, inspection of the spatial distribution, kinematics, and stellar population of the group II-b clusters provides good evidence that they may have originated from a satellite system. In this case, age and abundance can be decoupled so that they could be slightly younger than the genuine Galactic globular clusters of similar metallicity, in line with the predictions based on our HB models.

Figure 3, A to C, presents the spatial distribution of Galactic globular clusters (*12*), which shows that seven out of the nine group II-b clusters lie close to a single plane, making up a lengthy stream in the Galactocentric sky. They are NGC 5053, NGC 5466, M15, M30, M53, M68, and M92 (hereafter group II-b denotes only these seven clusters), with NGC 2419 and NGC 6426 being two exceptions. Because the plane defined by the clusters is nearly perpendicular to both the Galactic plane and to the *x* axis, no coordinate rotation has been applied to highlight the alignment of the clusters. It is intriguing that a certain cluster group selected by a stellar population has a spatial coherence. Moreover, the amount of spread in distances to the hypothetical plane [~±1 kiloparsecs (kpc)] coincides with the current uncertainty in distance determination for the clusters (*13*). Using

Monte Carlo models, we estimated the probability that seven out of nine halo objects at random, following the observed halo cluster density profile (*14*), may be located by chance within any 2 kpc–thick plane to be only 1.82%. A probability of chance alignment becomes less than 0.1% when considering the range in Galactocentric distance ($R_G$) of the aligned clusters, from 6.9 (M30) to 18.9 kpc (M53). If not by chance, the planar alignment and comparable $R_G$'s of group II-b clusters, together with their narrow range in metallicity, suggest that most of them were tidally pulled off from a satellite galaxy and, in turn, are following similar orbital trajectories.

To see if group II-b possesses a velocity coherence comparable to the spatial coherence discovered above, we plotted (Figs. 4, A and B) the space motion vectors of the group II-b clusters with the uncertainties in direction (*15*). Among the seven clusters of interest, six have published space velocities. M30, M53, and M92 are moving in a direction along the plane defined by group II-b to within the uncertainty (Fig. 4A). If one assumes an error on the order of 2.5 σ, NGC 5466 and M68 may also remain possible members of the proposed group II-b stream. The motion of M15, however, is roughly perpendicular to the group II-b stream. As seen from the Sun toward the Galactic center, M15, M30, M68, and M53 have a common counterclockwise motion (Fig. 4B). The velocities of M92 and NGC 5466 also show signs of following the general trend of counterclockwise motions shown by the rest of group II-b clusters. The space velocities show some indications of a systemic motion around the Galaxy. We have also investigated their kinematics with radial velocities alone (*12*) (Fig. 4C). A sinusoidal pattern marked by the group II-b clusters is consistent with a Keplerian motion of comparable magnitude and direction (*16, 17*).

To explore the possible association of group II-b with a satellite galaxy, we have examined the nearby companions of the Milky Way (*18*). Group II-b displays a spatial correlation with the well-known Magellanic Plane (*19, 20*) galaxies (Ursa Minor and Draco dwarf spheroidals, and the Magellanic Clouds) (Fig. 3D). Recently, a stellar stream in the halo of M31 (the Andromeda galaxy) was discovered (*21*), which lies along the satellites M32 and NGC 205 and is considered to be a result of interactions of either, or both, of the satellites with M31. The configuration of the Milky Way, the satellites in the Magellanic Plane and the stream of the group II-b clusters closely resembles that of M31, its satellites and the stream of stars. The fact that Ursa Minor, Draco, or the Small Magellanic Cloud presently contains at most only one globular cluster may argue against their direct association with group II-b. In this regard, the Large Magellanic Cloud (LMC), the Galaxy's largest satellite, is of interest because it is the only galaxy in the Magellanic Plane known to have a dozen old globular clusters in its own halo (*22*).

In terms of stellar population, recent *Hubble Space Telescope* observations reveal that the old LMC clusters are indistinguishable in age from the group II-b clusters (*23, 24*). Moreover, the clusters in the LMC's outer halo, where clusters could preferentially be stripped during the tidal disruption, have an HB type that is similar to that of group II-b for the corresponding metallicities (*25*). It is not surprising, therefore, that six of the seven LMC clusters with known $\langle P_{ab} \rangle$ and [Fe/H] (*22*) fall near the isochrone running through group II-b, including a cluster having $\langle P_{ab} \rangle$ and [Fe/H] values typical of group II-b (Fig. 1). Thus, the existence of odd Oosterhoff-intermediate clusters in the LMC argues in favor of their similarity to group II-b in stellar contents. From the viewpoint of kinematics, the proposed orbital plane of the Magellanic Clouds is perpendicular to the Galactic plane, and the sense of the orbits is counterclockwise as seen from the Sun toward the Galactic center (*26, 27*), which is in accordance with the distribution and motion of group II-b. Also, radial velocities of group II-b clusters are commensurate with those of clusters that were claimed to be associated with the Magellanic Clouds [Rup 106 and Pal 12 (*28*); IC 4499, Rup 106, Pal 4, Pal 12, and NGC 6101 (*17*); Pyxis (*29*); Pal 3, M3, and NGC 4147 (*11*)] and the Magellanic Plane galaxies (Fig. 4C).

The group II-b plane is about 5 kpc away from the Galactic center. A similar displacement is seen in the M31 stream, but the amount of the displacement in this group II-b case is relatively large, as compared to the apparent extent of their configuration. One possible explanation is that the offset could be due to the binary motion of the two Magellanic Clouds around each other; i.e., the center of mass of the Clouds would orbit the Galactic center. Recent dynamical studies (*26, 27*) suggest that the Clouds have been in a binary state for at least the past several billion years, with the typical separation of about 20 to 30 kpc (currently 21 kpc). With the current mass ratio of the Large and Small Clouds being about 5:1, the plane that is off-center to such a degree seems to fit in with the LMC hypothesis. Intuitively, the preservation of the arrangement of positions may imply that the capture occurred at a relatively recent epoch. Measurement of more accurate space velocities and detailed dynamical modeling will help to confirm the possible connection between group II-b and the LMC.

It has been pointed out that clusters of group I lie on retrograde orbits more frequently than do clusters of group II (*4*). Eight of the nine retrograde clusters are concentrated in group I (Fig. 1). Dividing clusters further into two groups according to $R_G$, we find that most retrograde clusters belong to the subpopulation of group I clusters having $R_G \geq 8$ kpc. Thus it appears that, apart from a few interlopers, most clusters with $R_G \geq 8$ kpc, such as the group I clusters on retrograde orbits and our group II-b clusters, show evidence for an accreted origin from satellite systems, following the younger isochrone (Fig. 1) (*4, 7*). We conclude that the Oosterhoff dichotomy and its close link with cluster metallicity, kinematics, and age have arisen as a result of two distinct mechanisms involved in the formation of the Galactic halo; the early formation of globular clusters in the proto-Galaxy (clusters in the inner halo and those in our group II-a) and the later accretion from satellite systems (most of the outer halo clusters including those in our group II-b).

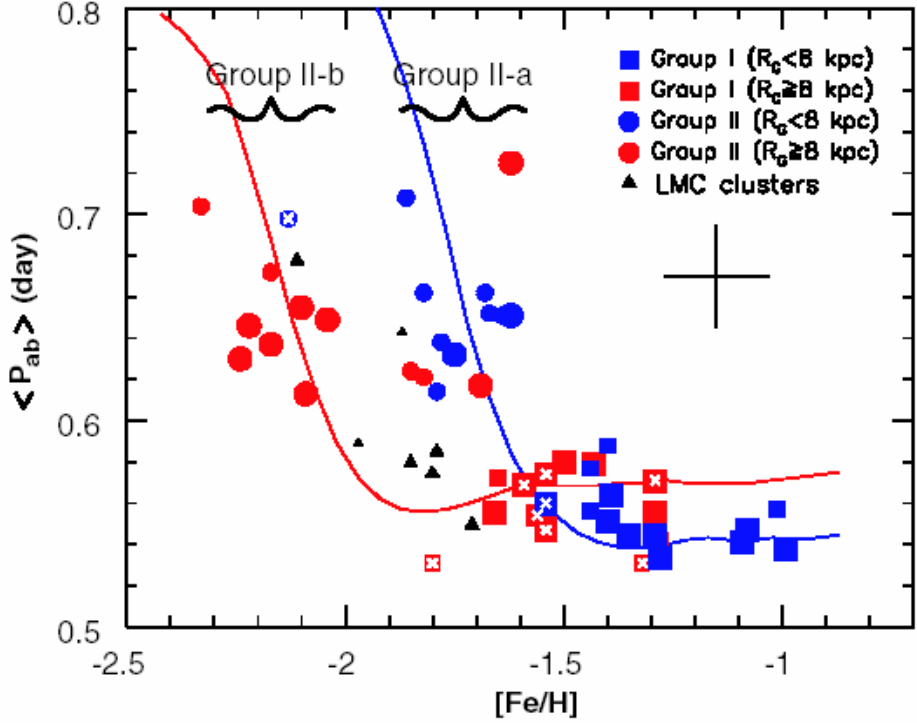

**Fig 1.** Correlation of the mean period of type *ab* RR Lyraes ($\langle P_{ab}\rangle$) (*2*) with cluster metallicity ([Fe/H]) (*12*) for globular clusters in the Galaxy and the LMC. The large (small) symbols are for clusters with 10 or more (3 or more but less than 10) type *ab* RR Lyraes. The typical error bar is shown. Clusters are divided into two distinct groups according to the mean period and the metal abundance; Oosterhoff group I (squares; $\langle P_{ab}\rangle \approx 0.55$ days, and [Fe/H]>-1.8) or group II (circles; $\langle P_{ab}\rangle \approx 0.65$ days, and [Fe/H]<-1.6). Clusters with $R_G$<8 kpc (blue symbols) are well reproduced by the models for the Δage=0-Gy population (blue isochrone), whereas most clusters with $R_G$≥8 kpc (red symbols) follow the model locus for slightly younger age (Δage=-1.2 Gy; red isochrone). The triangles represent the old globular clusters in the LMC (*22*). The crosses mark clusters on retrograde orbits (*4*).

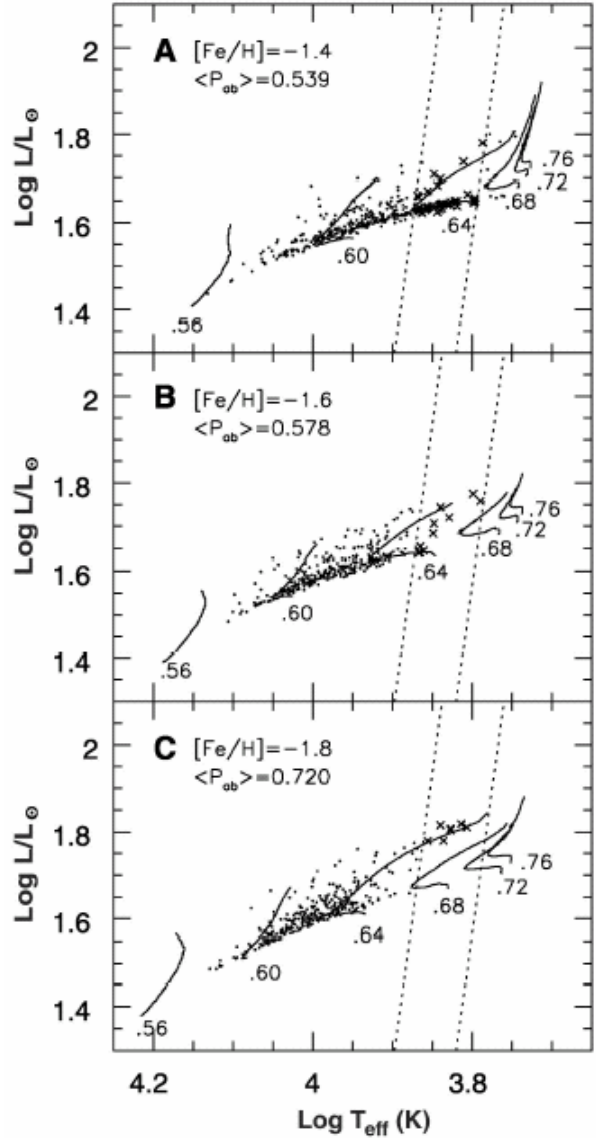

**Fig 2.** HB population models in the effective temperature (in kelvins) versus luminosity (in solar units) plane. Representative models are shown for [Fe/H]=-1.4 (**A**), [Fe/H]=-1.6 (**B**), and [Fe/H]=-1.8 (**C**). In each panel, [Fe/H] and $\langle P_{ab} \rangle$ in days are denoted, which correspond to the blue locus in Fig 1. Whereas HB stars outside the instability strip (shown by dotted lines) are normal HB stars (dots), stars within the strip are RR Lyrae variables (crosses) (*6*). The HB evolutionary tracks (*31*) are overlaid (shown by solid lines) and labeled by its total mass in solar units. The topology variation of the HB evolutionary track within the instability strip and the corresponding evolutionary effect from blue HB cause an abrupt rise in the mean RR Lyrae luminosity and $\langle P_{ab} \rangle$.

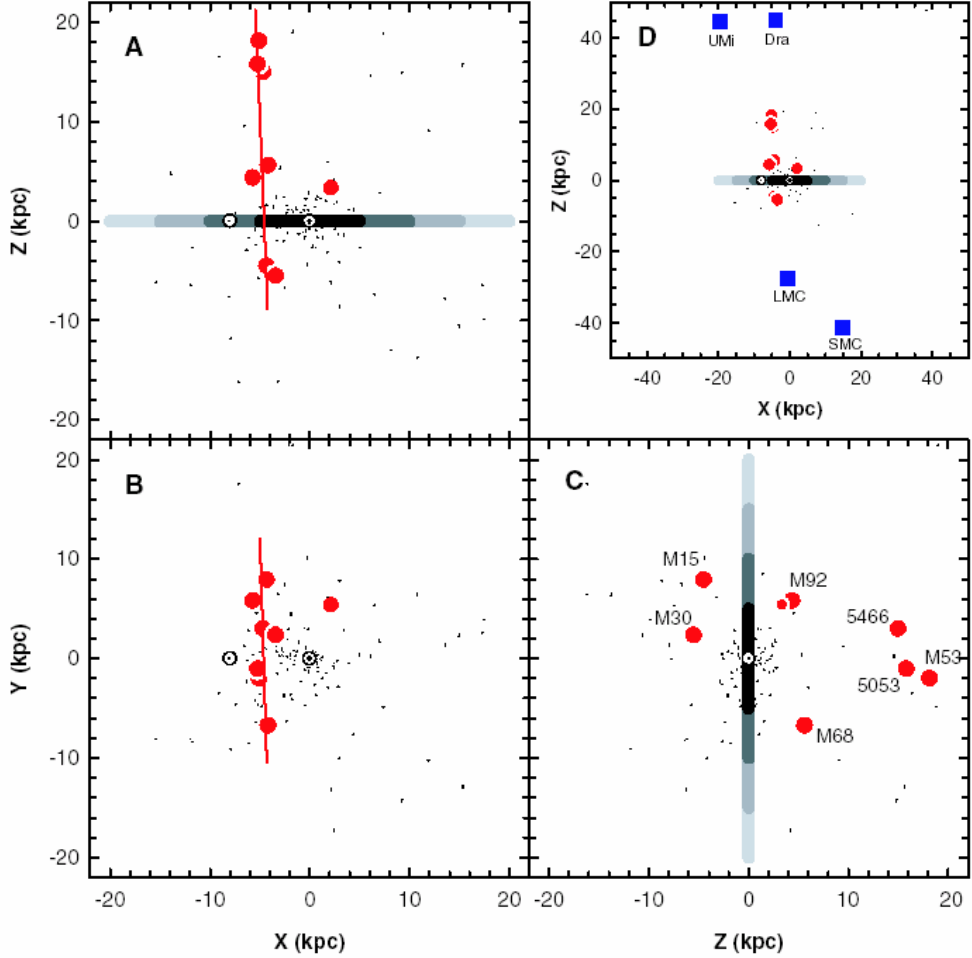

**Fig 3.** Spatial distribution of Galactic globular clusters (*12*) in a Cartesian coordinate system centered on the Galactic center. **(A)** Front view. **(B)** View from above the Galactic plane. **(C)** Side view. The *x* component is positive outward from the Galactic center (the circle with a plus), *y* is positive in the direction of Galactic rotation, and *z* is positive toward the north Galactic pole. The Galactic plane is drawn schematically, and the Sun (the circle with a dot) is at (-8, 0, 0) kpc. The large red circles are for the seven group II-b clusters that form a planar alignment, whereas the small red circle is NGC 6426 at (*x*, *y*, *z*)=(2.1, 5.4, 3.3) kpc. NGC 2419 at (-90.7, -0.6, 39.1) kpc is not shown. The name or NGC number of each group II-b cluster is labeled in panel (C). **(D)** Same as (A), but a 50 × 50 kpc window including the Magellanic Plane galaxies; LMC, Small Magellanic Cloud (SMC), and Draco (Dra) and Ursa Minor (UMi) dwarf spheroidal galaxies (blue squares) (*18*).

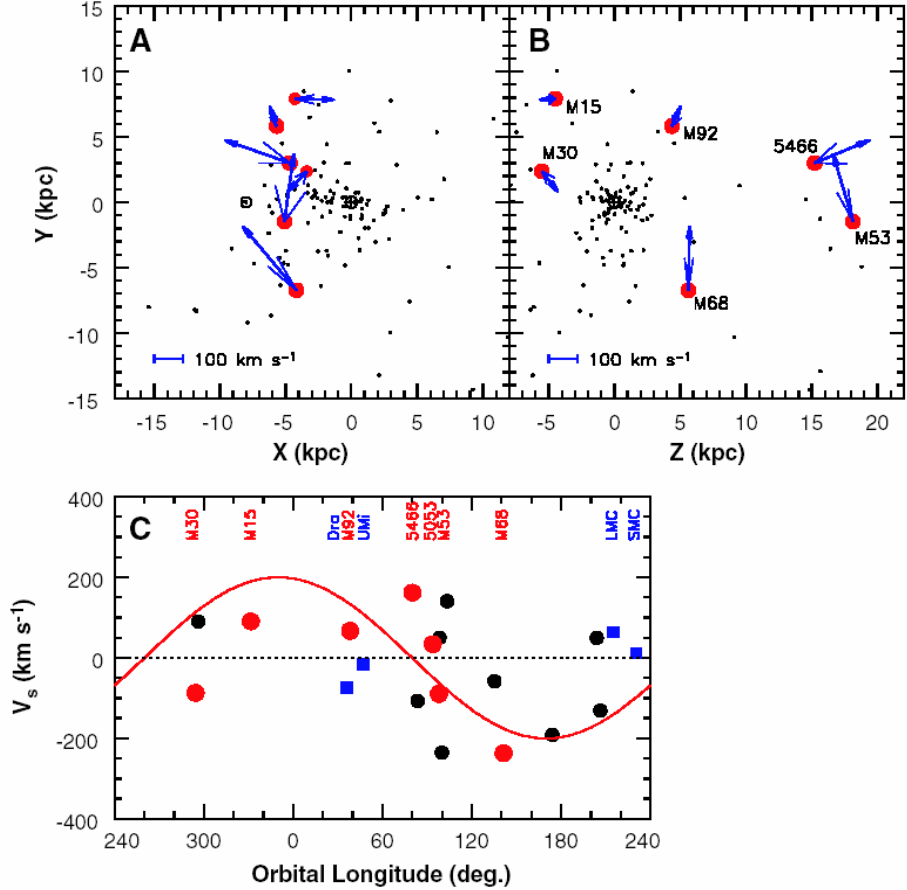

**Fig 4.** Kinematic analysis for the group II-b clusters. (**A**) Space motion vectors (thick arrows) of the group II-b clusters (red circles) with the uncertainties (thin lines) (*15*) are shown in the *x-y* plane (corresponding to Fig. 3B). The large and small red circles represent clusters with Z≥0 and Z<0 kpc, respectively. The Galactic center (the circle with a plus) and the Sun (the circle with a dot) are also marked. (**B**) Same as panel (A), but in the *z-y* plane (corresponding to Fig. 3C). (**C**) Radial velocity of a cluster observed at the Sun's position by an observer who is stationary with respect to the Galactic center ($V_s$) (*12*), is plotted against the orbital longitude measured along the plane defined by the group II-b clusters (*16*). Blue squares and black circles represent, respectively, the Magellanic Plane galaxies (which have their names along the edge) and globular clusters presumably associated with the Magellanic Clouds (see text).